\documentclass[twocolumn,showpacs,preprintnumbers,amsmath,amssymb]{revtex4-1}
\usepackage{graphicx}
\usepackage{dcolumn}
\usepackage{bm}
\usepackage{xcolor,soul}
\usepackage{multirow}
\newcommand{\beq}{\begin{equation}}
\newcommand{\eeq}{\end{equation}}
\newcommand{\beqa}{\begin{eqnarray}}
\newcommand{\eeqa}{\end{eqnarray}}
\newcommand{\rr}{{\cal R}}
\sethlcolor{yellow}
\setulcolor{red}
\usepackage{amsfonts}
\usepackage{amssymb}

\begin{document}

\title{Parallel ion strings in linear multipole traps}

\author{M. Marciante}
 \email{mathieu.marciante@etu.univ-provence.fr}
\author{C. Champenois}
\author{J. Pedregosa-Gutierrez}
\author{A. Calisti}
\author{M. Knoop}

\affiliation{Physique
des Interactions Ioniques et Mol\'eculaires, UMR 6633 CNRS et Aix-Marseille Universit\'e,
 Centre de Saint J\'er\^ome, Case C21,
13397 Marseille Cedex 20, France}%

\date{\today}

\begin{abstract}
Additional radio-frequency (rf) potentials applied to linear multipole traps create extra field nodes in the
radial plane which allow to confine single ions, or strings of ions, in  totally rf field-free regions. The
number of nodes depends on the order of the applied multipole potentials and their relative distance can be
easily tuned by the amplitude variation of the applied voltages. Simulations using molecular dynamics show that
strings of ions can be laser cooled down to the Doppler limit in all directions of space. Once cooled, organized
systems can be moved  with very limited heating, even if the cooling process is turned off.
\end{abstract}


\maketitle

Radio-frequency (rf) traps are very useful tools for a wide range of research interests such as high resolution
spectroscopy \cite{wolf09}, frequency standards \cite{chou10}, quantum information processing \cite{blatt08} and
quantum simulations \cite{johanning09}. The ability to trap and cool atomic or molecular species in a well
defined manner allows to  control and study the quantum behaviour of these systems. Most experiments require a
totally perturbation-free dynamics of the confined ions, and the micro-motion due to the rf trapping field can
sometimes be a disturbing factor. In these cases, cold ions are confined in a string configuration along the
longitudinal axis of a linear quadrupole trap, where the rf field vanishes.
For a number of applications, in particular in quantum logic and quantum simulations \cite{blatt08,johanning09},
it could be desirable to create more sites inside a trap where the ions do not undergo the rf parametric
excitation. Linear traps of higher order, here called multipole traps, offer larger regions of low rf electric
fields \cite{gerlich92} which allow to trap large samples with a reduced driven micro-motion, as compared to the
same sample trapped in a linear quadrupole trap \cite{prestage99}. Local minima can be induced in the trapping
potential by adding static voltages to these multipole rf voltages \cite{wester09,champenois10} but the ions
then undergo a rf driven motion which is detrimental when very low temperatures need to be reached. The
superimposition  of a lower order rf field onto the main trapping field generates  minima to the trapping
potential where the rf field is nulled. In such a configuration and depending on their number, the ions can
settle in each minimum as individual ions or as parallel strings of ions expanding in the axial direction. Due
to the absence of a local rf electric field, laser cooling can reduce the temperature of the ions as low as for
a chain of ions in a quadrupole trap.

In this letter we first give an example of principle of the proposed method by combining a quadrupole potential
with the potential created by  a linear octupole trap. To demonstrate how to reach the Doppler limit by Doppler
laser cooling we use molecular dynamics (m.d.) simulations of a 10-ion system, in a double line configuration.
In a second part, we discuss the general case of two combined rf potentials of different order. In these two
parts, we assume an ideal electric field for each superimposed rf field. In the third part of this letter, the
effect of a non-ideal geometry to generate the lowest order rf field  \cite{reuben96} is taken into account by
combining m.d. simulations with an analytic fit of the actual electric field. This analysis shows that,
providing  an extra  voltage which compensates for the   imperfections of the electric fields, the same
properties are observed and that our proposal still holds.

\begin{figure}[h]
  \centering
\includegraphics[width=3.cm]{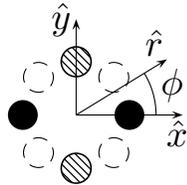}
  \caption{Connection of a quadrupole rf electric potential onto the electrodes of an octupolar trap. Black and hatched  rods show the different polarities of the applied $- V_4$ and $+ V_4$ potentials respectively, at time $t=0$ s. The broken line rods are only connected to the octupolar potential .}
  \label{fig_example}
\end{figure}

First, we describe the principle of the creation of extra rf field-free minima within the linear octupole trap
geometry. In order to create additional nodes in the rf electric field, a rf quadrupole potential, with
identical phase and frequency as the octupole potential, is superimposed onto the former.
By the superposition theorem, the resulting rf electric potential $\Phi_{\text{rf}}$ in the radial plane is
given by
\beq
    \Phi_{\text{rf}}= \Phi_{\text{8rf}}+\Phi_{\text{4rf}} ,
\eeq
where the expression of the $2k$-pole electric potentials $\Phi_{(2k)\text{rf}}$ using the polar coordinates
(see Fig.~\ref{fig_example}) can be approximated by \cite{gerlich92}:
\beq
    \Phi_{(2k)\text{rf}}\left(\rr,\phi,t\right) = -V_{2k} \rr^k \cos\left(k\phi\right) \cos\left(\Omega t\right)
\eeq
with $\Omega$ and $V_{2k}$ respectively the frequency and the  amplitude of the applied rf potentials, and $\rr
= r/r_0$ the relative distance to the trap center, scaled to the inner radius of the trap $r_0$. The components
of the electric field $\vec E = - \vec\nabla \Phi_{\text{rf}}$ are:
\beqa
    E_r & = &+\frac{2}{r_0} \left[2 V_8 \rr^3 \cos\left(4\phi\right)
        + V_4 \rr \cos\left(2\phi\right) \right]\cos\left(\Omega t\right), \label{eq_Er}\\
    E_{\phi} & = & -\frac{2}{r_0} \left[2 V_8 \rr^3 \sin\left(4\phi\right)
        + V_4 \rr \sin\left(2\phi\right)\right]\cos\left(\Omega t\right). \label{eq_Ephi}
\eeqa
With our choice of initial phase and frame reference (sketched on Fig.~\ref{fig_example}), it is straightforward
from  Eq.~\ref{eq_Ephi} that  $E_{\phi}$ can be cancelled for any direction defined by $\phi_n = n\pi/2$, with
$n \in \mathbb{Z}$.
The relative variation of the cosines in Eq.~\ref{eq_Er} makes then possible a total rf field cancellation at
the relative radial positions $\rr_{\text{0f}}$ for $\phi_{\pm1}=\pm \pi/2$ (the $y$-axis in our case, see
Fig.~\ref{fig_example}). This relative radial position depends on the ratio of the two rf voltages as :
\beq
    \rr_{\text{0f}}^2 = V_4/2V_8.
    \label{eq_r0f}
\eeq
In the adiabatic approximation, where the rf period is much smaller than the time scale of the motion induced by
the spatial variation  of the electric field amplitude (the so-called macro-motion) \cite{dehmelt67}, the
dynamics of an ion of charge $q_e$ and mass $m$ inside the trap can be understood using  the  pseudopotential
(psp) $\Psi$, which results from the time-average of the rf electric field $\Psi =q_e^2 \overline{\|\vec
E\|^2}/2 m \Omega^2$, where the overline holds for the average over one rf period. The psp does not obey the
superposition theorem and in our case of superimposed rf fields, the resulting psp writes as :
\beq
    \Psi\left(\rr,\phi\right)\ =\ \Psi_{8}(\rr) +\Psi_{4}(\rr) +2 \sqrt{\Psi_{8}(\rr) \Psi_{4}(\rr)}\ \cos\left(2\phi\right)
    \label{eq_Psi}
\eeq
where the $\Psi_{2k}(\rr)$ are the usual $2k$-polar psp, $2k$ being the number of electrodes \cite{gerlich92}:
\beq
    \Psi_{2k}\left(\rr\right) = \frac{k^2}{4} \frac{q_e^2V_{2k}^2}{m\ r_0^2\ \Omega^2}\ \rr^{2k-2}= \psi_{2k}\ \rr^{2k-2}.
 \label{eq_Psi2k}
\eeq
By definition of the psp, the pseudo-potential wells exactly match the nodes of the rf electric field.
Eq.~\ref{eq_Psi} shows that two extra psp minima, compared to the simple $\Psi_8$ case, are expected for $\rr
\neq0$ if $\cos(2\phi)=-1$.

However, the total effective trapping potential $U_{\text{trp}}$ responsible for the ion confinement results
from the psp itself and the  static potential required to trap along the symmetry axis  $Oz$ and which can be
approximated by a harmonic confinement $m\omega_z^2 z^2/2$. To obey the Laplace equation, this trapping
potential has a de-confining  counterpart $-m\omega_z^2 r^2/4$ which must be taken into account \cite{drewsen00,
champenois10}. In the case of an extra static octupole potential $V_{\text{st}}$ applied to the electrodes, the
effective trapping potential now reads
\beq
    U_{\text{trp}} = \Psi + q_eV_{\text{st}}\rr^4 \cos\left(4\phi\right)
                    + \frac{1}{2}\ m\ \omega_z^2 \left(z^2-\frac{\rr^2}{2}r_0^2\right)
    \label{eq_Hamilton}
\eeq
To make sure that the total potential minima still match  the rf field-free positions, it is mandatory to
compensate the radial de-confining force by an appropriate choice of the static potential $V_{\text{st}}$. The
matching of these two forces at the field free positions ($\rr_{\text{0f}}, \phi_{\pm1})$ is obtained under the
condition:
\beq
    q_eV_{\text{st}} = m\ \omega_z^2\ r_0^2\ /\ 8\rr_{\text{0f}}^2
    \label{eq_condition}
\eeq
To remain as general as possible, we have considered the extra octupole static voltage to compensate for the
deconfining effect of the axial confinement. The use of a  quadrupole voltage leads to a similar condition which
theoretically does not depend on $\rr_{\text{0f}}$ (cf. discussion of the realization below). We would like to
point out that the rf field cancellation condition Eq.~\ref{eq_r0f} as well as the matching condition
Eq.~\ref{eq_condition} do not depend on the mass of the trapped particles and the process described here can
apply to a multi-species system (because $ \omega_z^2 \propto 1/m$).

The lowest order expansion of the effective trapping potential around its minima ($\rr_{\text{0f}},
\phi_{\pm1})$ allows the definition of characteristic frequencies for the motion in the radial and angular
directions, if Eq.~\ref{eq_condition} is satisfied :
\beqa
    \omega_r^2 &=& \omega_z^2\ \left[\frac{8 \psi_4}{m\ r_0^2\ \omega_z^2}\ +1\right]=4\omega_u^2+\omega_z^2,
    \label{eq_omegaR} \\
    \omega_\phi^2 &=& \omega_z^2\ \left[\frac{8 \psi_4}{m\ r_0^2\ \omega_z^2}\ -2\right]=4\omega_u^2-2\omega_z^2,
    \label{eq_omegaPhi}
\eeqa
where $\omega_u$ is the  frequency of  motion in the radial plane, when only the rf quadrupole potential is
applied. The local steepness of the potential, given in Eq.~\ref{eq_omegaR} and \ref{eq_omegaPhi}, is controlled
by the quadrupole  potential $V_4$ and the axial confinement $\omega_z$, thus ruling the morphology of a cold
sample in the local 3D anisotropic harmonic potential $(\omega_r, \omega_{\phi}, \omega_z)$
\cite{schiffer93,dubin93,marciante10}. A close look at Eq.~\ref{eq_omegaPhi} shows that it takes $q_e V_4 >
mr_0^2 \omega_z \Omega /2$ to reach this local 3D harmonic confinement. The local potential being independent of
the octupole potential $V_8$, this last parameter can be tuned to  independently control the distance between
the two minima.

\begin{figure}[h]
  \centering
  \begin{tabular}{c}
  \hspace{0.20cm}
  \includegraphics[height=2cm,trim=0cm 0cm 0cm 0cm,clip]{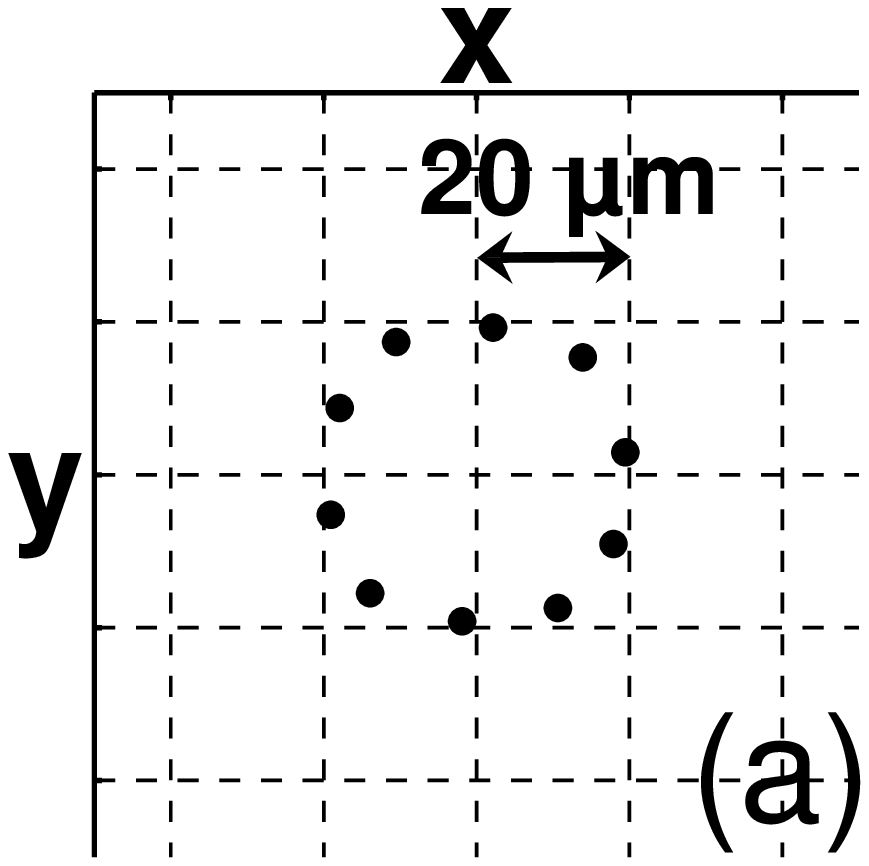}
  \hspace{0.20cm}
  \includegraphics[height=2cm,trim=0cm 0cm 0cm 0cm,clip]{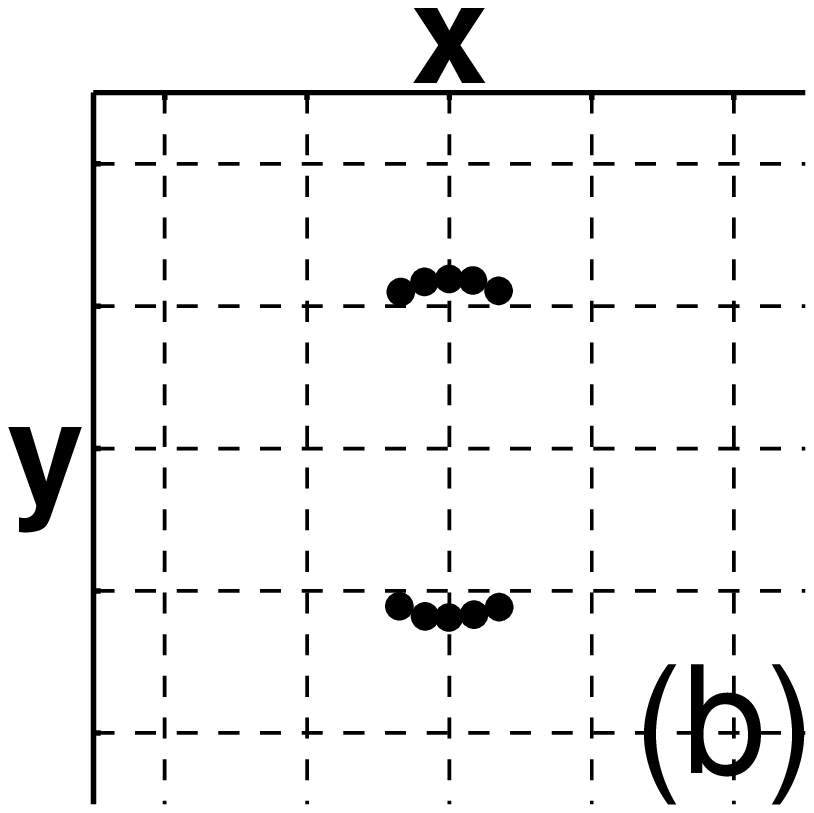}
  \hspace{0.15cm}
  \includegraphics[height=2cm,trim=0cm 0cm 0.1cm 0cm,clip]{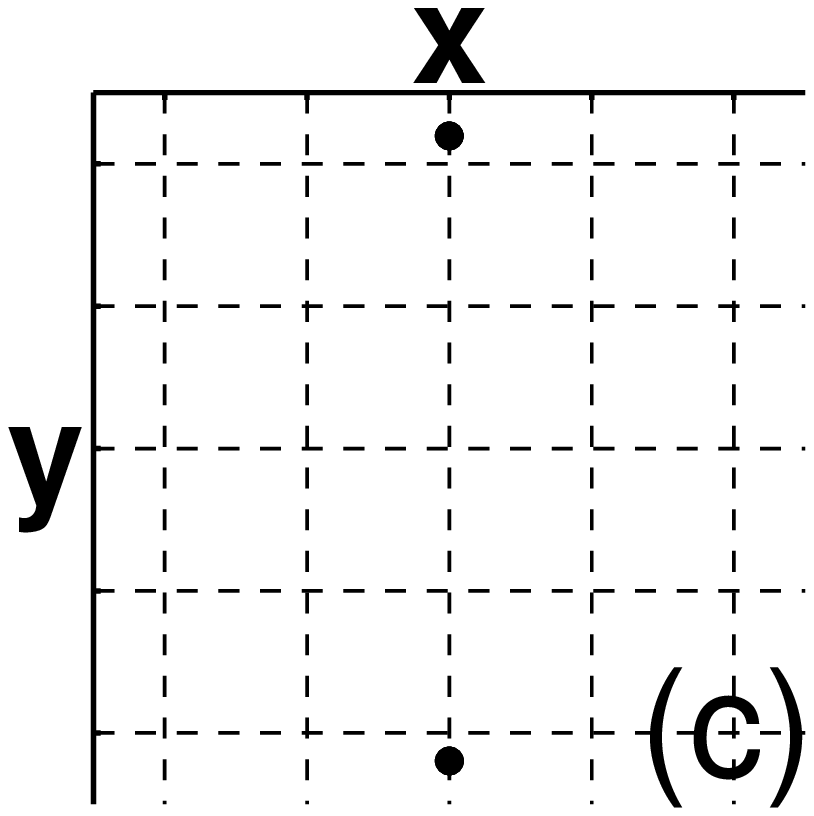}
  \includegraphics[height=2cm,trim=0.2cm 0.15cm 0cm -0.2cm,clip]{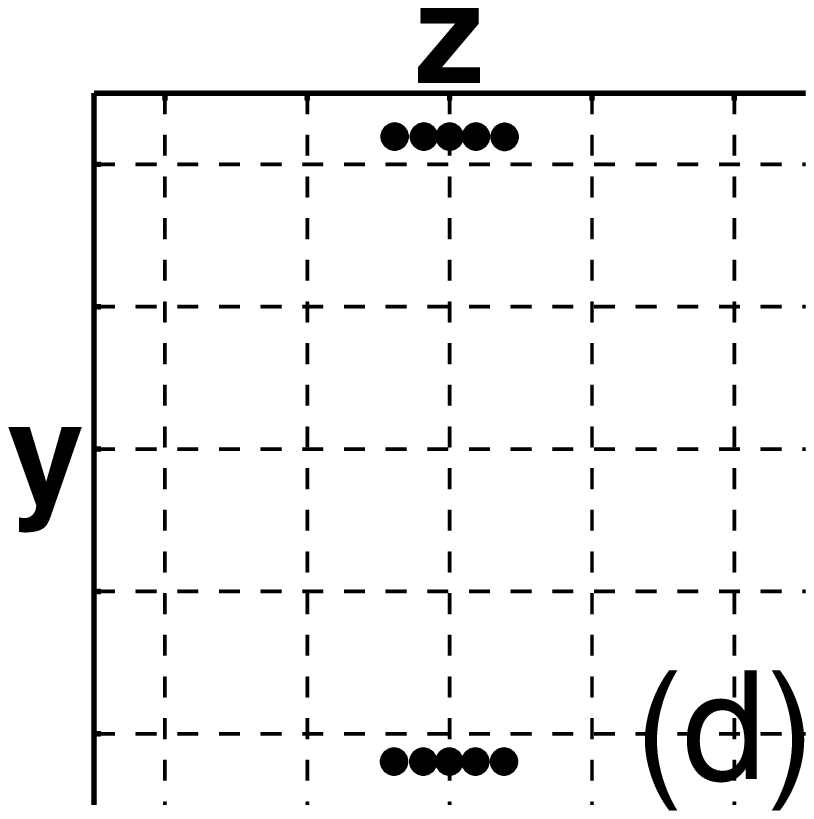} \\
  \includegraphics[width=8cm,trim=3.cm 0.0cm 6.0cm 1.0cm,clip]{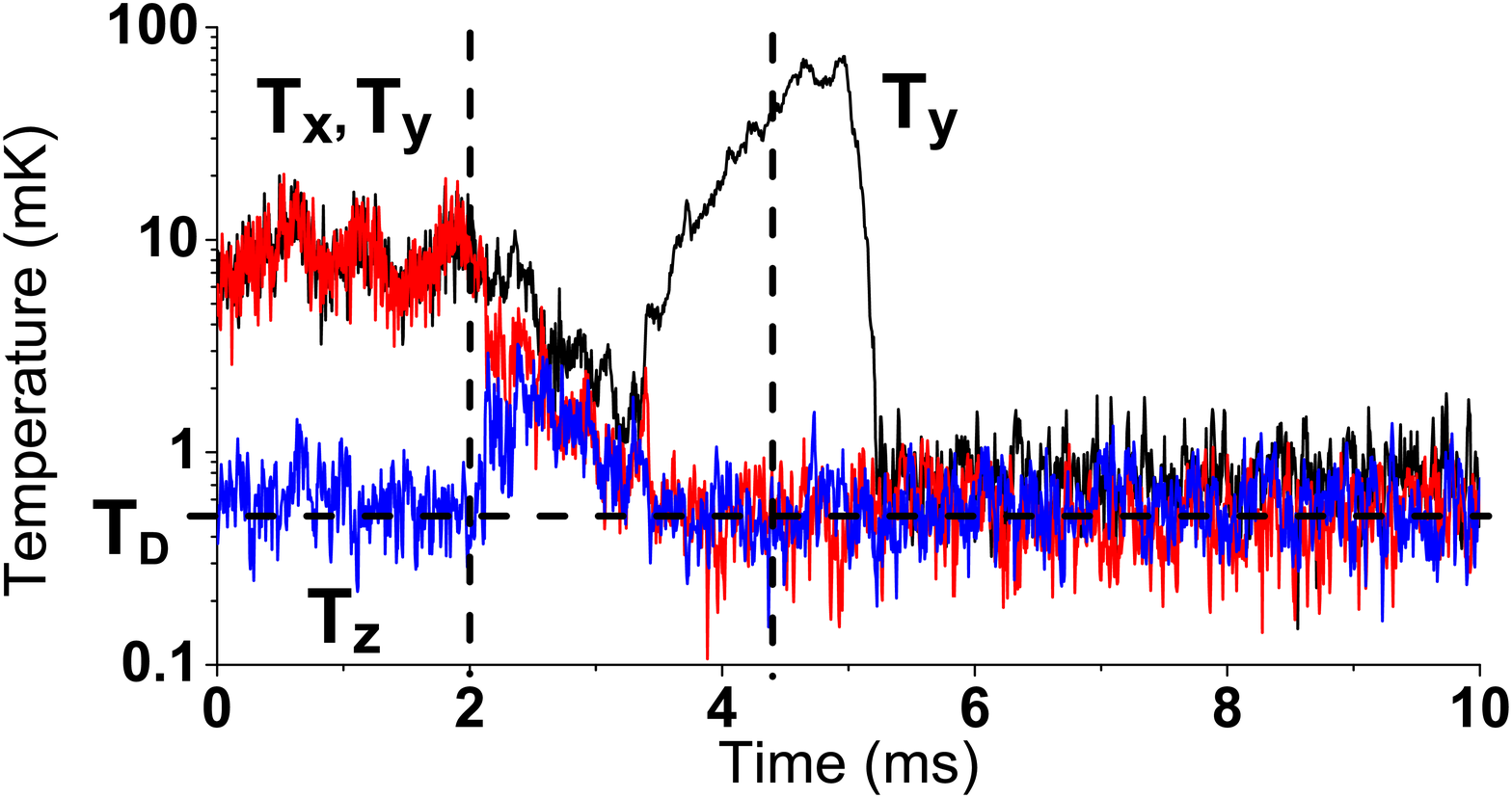}
  \end{tabular}
  \caption{(Color online) (a)-(d) Pictures of the ion structure at 3 different times
  (roughly corresponding to the temperature graph below).
  (e) Temperatures of the different directions of motion for a simulated dynamics
  detailed in the text. The horizontal dashed line shows the Doppler limit temperature for calcium ions and
  the vertical dashed lines   materialize first, when the quadrupole potential is switched on and second, when
  the static octupole potential is switched on.    }
  \label{fig_temperatures}
\end{figure}

In a second step, we show that ions can effectively be cooled to the Doppler limit and localized in the two
extra minima in a controlled manner. For this purpose we carry out m.d. simulations of a set of 10 calcium ions,
Doppler laser-cooled along the three directions of space in an octupole trap. The m.d procedure for laser
cooling takes into account the  absorption and emission processes of photons on the involved atomic transitions.
Details of its modeling as well as the definition of temperature used are explained in \cite{marciante10}. The
rf electric fields obey Eq.~\ref{eq_Er} and \ref{eq_Ephi} and at $t=0$, the rf quadrupole potential $V_4$ and
the static octupole potential $V_{\text{st}}$ are off. The amplitude of the rf octupole potential $V_8$ is such
that cold ions form a $\sim 20$ $\mu$m radius ring in the $z=0$ plane for $\omega_z=2\pi \times 1$~MHz, shown on
figure~\ref{fig_temperatures}a. The stability of such a structure has been studied
elsewhere \cite{okada07,champenois10}, where it is shown that the Doppler limit can be reached for the motion
along the longitudinal direction and that the cooling of the motion in the radial plane is limited by rf heating
\cite{champenois10}. At $t=2$~ms, the rf amplitude $V_4$ is switched on, breaking the cylindrical symmetry of
the trapping potential and producing the separation of the ring in two sets of five ions, as shown on
Fig.~\ref{fig_temperatures}b.
From $t=2$ ms to $t=4.4$ ms, the amplitude $V_4$ is linearly increased, increasing the distance between the two
sets of ions and the steepness of the local potential to reach a separation of $0.22 r_0$. For $t=4.4$ ms, the
ions form two strings along the z-axis (Fig.~\ref{fig_temperatures}c and Fig.~\ref{fig_temperatures}d). At this time,
$V_4$ is kept unchanged and $V_{\text{st}}$ is switched on from 0 to the value fulfilling Eq.~\ref{eq_condition}
(1.7 V) in 1~ms,  matching  the equilibrium positions of the strings of ions and the field-free positions  and
reaching characteristic frequencies $\omega_r=2\pi \times 4.2$~MHz and $\omega_{\phi}=2\pi \times 3.8$~MHz. The
signature of this position matching is the abrupt drop of $T_y$, the  temperature  of the motion along $Oy$,
from 0.1~K to the Doppler limit (see Fig.~\ref{fig_temperatures}e).

Another signature of this position matching and of the interest of this new ion  organization is the lack of
rf-heating observed in the simulations when the cooling lasers are switched off, once the parallel lines are
formed. Furthermore, tuning the amplitude $V_{8}$ and the static potential $V_{\text{st}}$ accordingly, the ion
strings can be moved along the $Oy$ axis. This relative position tuning can be performed with very limited
heating as long as the Coulomb repulsion between the two chains is  negligible. When it is not the case, this
repulsion has to be incorporated in the force compensation condition and a larger static voltage than the one
given in Eq.~\ref{eq_condition} is required.

\begin{figure}[h]
  \centering
  \includegraphics[width=7cm]{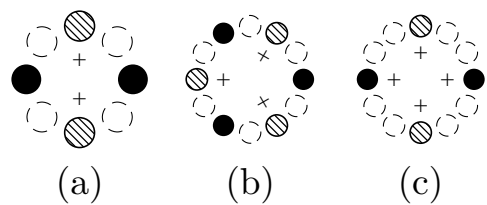}
   \caption{Diagram of the rf cylindrical rods of different traps where the filled and hatched rods are,
   respectively, the additional $-V_{2p}$ and $+V_{2p}$ at $t=0$. Inner crosses stand for the extra field-free
   positions of (a) octupole trap with additional quadrupole potential, (b) dodecapole trap with hexapole
   potential, and (c) dodecapole trap with quadrupole potential.}
  \label{fig_combine}
\end{figure}

It is possible to generalize the concept introduced previously. We now assume a $2k$-pole linear trap on which
an rf potential $V_{2k}$ is applied. An additional  $V_{2p}$ rf potential with identical frequency and phase is
superimposed on $2p$ electrodes to generate a rf contribution with a $p$-order of rotational symmetry. Each rf
contribution builds its own psp $\Psi_{2k}$ and $\Psi_{2p}$ (see Eq.~\ref{eq_Psi2k}) and the resulting psp can
be written as :
\beq
    \Psi = \Psi_{2k} +\Psi_{2p} +2 \sqrt{\Psi_{2k}\Psi_{2p}}\ \cos\left(\left(k-p\right)\phi\right).
\eeq
The extra  psp minima created by cancellation of the rf electric fields are located where
$\phi_n=(2n+1)\pi/(k-p)$ (for $\cos((k-p)\phi)=-1$) and   $\rr=\rr_{\text{0f}}$ such that
\beq
    \rr_{\text{0f}}^{k-p} = pV_{2p}/kV_{2k} .
\eeq
To compensate for the deconfining contribution of the axial confinement at the psp minima, an additional  static
potential with a $2\pi/(k-p)$ rotational symmetry is required to produce the same confining contribution along
all the $\phi_n=(2n+1)\pi/(k-p)$ lines. Depending on the choice of $k$ and $p$, this extra static potential may
have a different order of symmetry but a cancellation condition equivalent to Eq.~\ref{eq_condition} can always
be found. As an example, four parallel strings can be formed by the superimposition of an rf quadrupole
potential to an rf dodecapole potential, their location match the rf-field nodes providing  a static voltage
with an octupole symmetry is applied (see figure~\ref{fig_combine} for different configurations).
The order of symmetry $s$ of the static potential $V_{st}$ must be a multiple of $(k-p)$. For  the case
introduced in the first part of this letter, we used $s=2(k-p)=4$. This choice has an influence on the
characteristic frequencies of the motion of the ions in the extra minima. Indeed, a lowest order expansion of
the total trapping potential around these positions results in frequency expressions which are generalizations
of Eq.~\ref{eq_omegaR} and \ref{eq_omegaPhi} :
\beqa
    \omega_r^2 &=& 2\left(k-p\right)^2 \frac{\psi_{2p}}{m\ r_0^2}\ \rr_{\text{0f}}^{2p-4}
    +\left(s-2\right)\ \frac{\omega_z^2}{2},
    \label{eq_omegaRs} \\
    \omega_\phi^2 &=& 2\left(k-p\right)^2 \frac{\psi_{2p}}{m\ r_0^2}\ \rr_{\text{0f}}^{2p-4}
    -s\ \frac{\omega_z^2}{2}.
    \label{eq_omegaPhis}
\eeqa
The choice of  an order $s$ higher than strictly required by the rotational symmetry  may then be justified by
the need for high frequencies $\omega_r$ compared to $\omega_{\phi}$, which can be used to control the
morphology of the subsystems (string, zig-zag...) , at the expense of the static voltage itself, which scales as
$1/(s \rr_{\text{0f}}^{s-2})$.

\begin{figure}
  \centering
  \includegraphics[width=8.5cm]{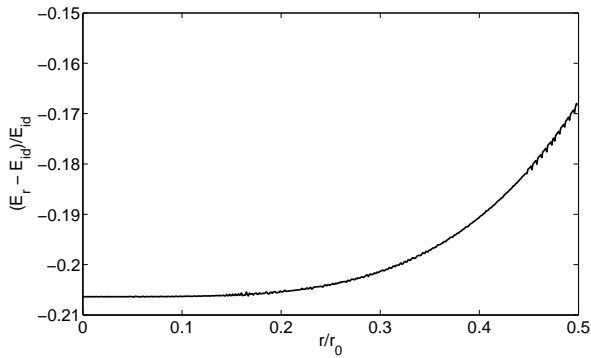}
   \caption{Normalized difference of the quadrupole electric field, $E_r$, and the ideal one, $E_{id}$, along
   the radial direction under the angle $\phi_{\pm1}$.}
  \label{fig_deviation}
\end{figure}

Until now, we have assumed equations for the rf electric field and its psp which correspond to the ideal case of
perfectly machined electrodes. Even if the trap design is as ideal as possible, it can not be ideal for both the
highest and the lowest order potential at the same time. To illustrate such an effect, let us go back to our
first example of an extra quadrupole potential applied to the octupole trap electrodes and let's assume the
linear octupole trap has round electrodes designed to generate an rf electric field as close as possible as the
ideal equation \cite{konenkov10}. On the quadrupole point of view, these electrodes are too small to produce an
ideal quadrupole field inside the trap \cite{reuben96} and most importantly, the four non-connected electrodes
are equivalent to grounded rods which induce a large deviation from the ideal rf field. Using  SIMION software
\cite{simion}, we have calculated the potential created by a quadrupole-like voltage connection on the octupole
trap used for the previous m.d. simulations. Its relevant characteristics are an inner radius of $r_0=0.2$~mm
and an electrode radius of $r_0/3$, which corresponds to the optimisation of the octupole potential
\cite{konenkov10}.
The calculation shows that the actual quadrupole potential deviates from the ideal one but keeps the quadrupole
symmetry.
In figure \ref{fig_deviation} is drawn the normalized difference of the actual quadrupole field component $E_r$
with respect to the ideal one, $E_{id}$, along the radial direction under the angle $\phi_{\pm1}$. This value is
constant in the central part of the trap ($r<r_0/5$) where, providing an $E_r$ amplitude 20\% higher, the fields can
be considered identical. As an example, the ion ring in Figure
\ref{fig_temperatures} fills a tenth of the actual trap dimensions.
Figure \ref{fig_deviation} also shows that Eq.~\ref{eq_condition} becomes position-dependent for an extra
quadrupole static voltage, and thus justifies the use of the optimized static octupole voltage if more distant
positions were required.

The m.d. simulations using the calculated potential, fitted by a polynomial equation \cite{pedregosa10a} show
that the same procedure to create two strings, to bring them to the rf nodes and to laser cool them to the
Doppler limit can be performed.

We have shown that a superimposed lower-order rf potential onto the main trapping one adds supplementary nodes
in the rf field. An ensemble of ions can be separated in several sub-sets in a controlled manner  and
laser-cooled to the Doppler limit. Once cooled, they can be considered as in a 3D anisotropic harmonic potential
and controlled like in a linear quadrupole trap.  This opens the way to create parallel ion strings with a large
choice of  geometries.
The distance between the potential minima can be continuously tuned to reach conditions where the Coulomb
repulsion between the subsets implies a strong enough correlation to influence their relative equilibrium
positions.
The different patterns formed by neighboring ions seem specifically interesting for quantum simulations
\cite{buluta09,brennen09}, but may certainly find interesting applications in other domains.


%

\end{document}